\documentclass[twocolumn,prb]{revtex4}
\usepackage{graphicx}    
\usepackage{dcolumn}     
\usepackage{bm}          

\begin{document}

\title{A quantum Monte Carlo study on the 
superconducting Kosterlitz-Thouless transition 
of the attractive Hubbard model on a triangular lattice}
\author{Tsuguhito Nakano}
\author{Kazuhiko Kuroki}
\affiliation{Department of Applied Physics and
Chemistry, The University of Electro-Communications,
Chofu, Tokyo 182-8585, Japan}

\date{\today}
\begin{abstract}
We study the superconducting Kosterlitz-Thouless transition 
of the attractive Hubbard model on a two-dimensional triangular 
lattice using auxiliary field quantum Monte Carlo method for 
system sizes up to $12\times 12$ sites. Combining three methods 
to analyze the numerical data, we find, for the attractive interaction
 of $U=-4t$, that the transition temperature stays almost constant
 within the band filling range of $1.0 < n < 1.4$, while it is found to
 be much lower in the $n<1$ region.
\end{abstract}
\maketitle
\section{Introduction}

Discovery of superconductivity in layered 
materials or quasi-two-dimensional systems in the 
past several decades has brought up great interest in the 
physics of low dimensional superconductors. Among those are 
the high $T_c$ cuprates,\cite{BM} a ruthenate Sr$_2$RuO$_4$,\cite{Maeno}
a cobaltate Na$_x$CoO$_2\cdot y$ H$_2$O,\cite{Takada}
MgB$_2$,\cite{Akimitsu} a heavy electron system
CeCoIn$_5$,\cite{Petrovic} and organic conductors such as
(BEDT-TTF)$_2$X (X is an anion).\cite{McKenzie}

Theoretically, it is well known by Mermin-Wagner's theorem that 
no off-diagonal long range order takes place at finite temperature 
in purely two-dimensional(2D) systems.\cite{MW} In the case of pure 2D, 
superconducting transition is expected to be of the 
Kosterlitz-Thouless (KT) type.\cite{KT} The superconducting KT transition has 
previously been studied using finite temperature 
auxiliary field quantum Monte Carlo (AFQMC) technique for the 
attractive Hubbard model, that is, the Hubbard model with 
a negative on-site $U$, on a square 
lattice\cite{MS} and on a triangular lattice.\cite{Santos} 
Nowadays, a renewed interest for the KT transition on triangular
lattices has arisen because unconventional superconductivity has been observed in
materials having (anisotropic) triangular lattice structure such as
Na$_x$CoO$_2\cdot y$H$_2$O and (BEDT-TTF)$_2$X.

For the square lattice, it has been shown that the KT superconducting 
transition temperature $T_0
\sim 0.1t$, $t$ is the hopping integral, for square lattice with $n=0.87$ ($n$ is the band filling)
away from the half filling $n=1.0$, where charge density wave (CDW)
ordering takes place.\cite{MS} As for the triangular lattice, it has
been concluded that $T_0$ takes its maximum at around half-filling due
to the absence of CDW ordering, while $T_0$ is low near $n=1.4$,where
the density of states becomes large due to the van Hove singularity at
$n=1.5$. 

Since previous studies have been restricted to small system sizes,
here we revisit the problem of the superconducting KT transition of 
the attractive Hubbard model on a triangular lattice using AFQMC technique. 
Combining three methods of analyzing the numerical data, 
we find that the KT transition temperature 
stays almost constant within the band filling region of $1<n<1.4$.Our
results suggest that the estimation of the KT transition temperature
using finite size scaling requires some caution when the density of
states near the Fermi level is small and the calculation is restricted
to small system sizes.


\section{Formulation}
\subsection{Model}
The Hamiltonian of the attractive Hubbard model is given 
in standard notation as 
\begin{eqnarray}
{\cal H} &=&
-t \sum_{\left< i,j \right>,\sigma}(c_{i,\sigma}^{\dagger}c_{j,\sigma}+c_{j,\sigma}^{\dagger}c_{i,\sigma}) \nonumber \\
&&+U \sum_i n_{i,\uparrow}n_{i,\downarrow} - \mu \sum_{i,\sigma}n_{i,\sigma},
\end{eqnarray}
where $c_{i,\sigma}^{\dagger}(c_{i,\sigma})$ is a fermion creation
(annihilation) operator at site $i$ with spin $\sigma$, and
$n_{i,\sigma}=c_{i,\sigma}^{\dagger}c_{i,\sigma}$,$<i,j>$ denotes a pair
of nearest neighbors on square or triangular lattice having periodic
boundary condition.The chemical potential $\mu$ controls the band filling
$n$ (average number of electrons per site).  
The hopping parameter $t$ is taken as the unit of the  energy
and is set equal to unity throughout the study.
As for the lattice structure, we mainly concentrate on the 
triangular lattice, but we also perform calculation on the square 
lattice at the band filling of $n=0.87$ in order to make comparison 
with previous studies and with the triangular lattice. 
The on-site attraction $U$ is fixed at $U/t=-4.0$ on both lattices, which 
also enables us to make comparison with the previous studies.

\subsection{Pairing Correlation Function}

Above the superconducting KT transition temperature $T_0$, 
the on-site ($s$-wave) pairing correlation function 
decays exponentially, 
\begin{equation}
 <c_{i,\uparrow}^{\dagger}c_{i,\downarrow}^{\dagger}c_{j,\uparrow}c_{j,\downarrow}> \sim \exp(-r/\xi) 
\label{eqrawcor0}.
\end{equation}

By contrast, the pairing correlation function exhibits a power-law decay
\begin{equation}
 <c_{i,\uparrow}^{\dagger}c_{i,\downarrow}^{\dagger}c_{j,\downarrow}c_{j,\uparrow}> \sim r^{-\eta} \label{eqrawcor},
\end{equation}
where $\eta=0.25$ for $T\rightarrow T_0$. 
In the actual AFQMC calculation, we calculate a summation of the 
pairing correlation function, $P_s$, given as 
\begin{eqnarray}
 P_s &=& \left<\Delta^{\dagger} \Delta + \Delta \Delta^{\dagger}\right> \label{eqPs} \\
\Delta^{\dagger} &=& \frac{1}{\sqrt{N}}\sum_i c_{i,\uparrow}^{\dagger}c_{i,\downarrow}^{\dagger}
\end{eqnarray}
Due to the decaying behavior of the pairing correlation function
mentioned above, 
$P_s$ increases as $N$ is increased below $T_0$, while it remains
constant above $T_0$.



In the present study, finite temperature AFQMC is used to calculate $P_s$.
\cite{Hirsch,White}
The number of Trotter decomposition slices $L$ is chosen to 
satisfy the condition $L \ge 12\beta$ ($\beta=1/T$), so that  
$\Delta \tau\le 0.084/t$ is satisfied, where $\Delta \tau=\beta/L$.
$3000-30000$ Monte Carlo sweeps, depending on the temperature and 
system size, have been taken to assure sufficiently small 
statistical errors.
We have performed calculation on system sizes from $4^2$ to $12^2$ 
sites.

In order to obtain $T_0$ from the AFQMC data of $P_s$,
we use three methods, two of which are based on finite
size scaling, while the other one is a more straightforward method.

\subsection{Finite size scaling}

As shown in eq.{\ref{eqrawcor}}, the pairing correlation decays as
$r^{-\eta}$ below $T_0$ for a large system size. Hence $P_s$ is proportional to ${N_x}^{2-\eta}$ with system size $N=N_x^2$. In a finite size system,scaling variable $N_x/\xi$ ,where  $\xi$ is the correlation length, becomes more 
important.Taking this into account,the scaling hypothesis assumes the 
following behavior for $P_s$.
\begin{eqnarray}
P_s(T,N_x) = N_x^{2-\eta }f(N_x/\xi) \label{eqFSS}\\
\xi = \exp(A/(T-T_0)^{1/2}) \label{eqxi},
\end{eqnarray}
where $A$ is a constant and $f(x)$ is a certain scaling function. 
Since $\xi \rightarrow \infty$ for $T\rightarrow T_0$, 
$f(x)=f(0)$ regardless of the system size $N$ at $T=T_0$. 
Thus, if we plot $P_sN_x^{-2+\eta}$ as functions of 
$T$(or $\beta$) for different system sizes, 
they should coincide regardless of the system size at $T=T_0$.
\cite{Santos}

Another method to identify $T_0$ is to plot $P_sN_x^{-2+\eta}$ as a function of
$N_x/\xi$, where $T_0$ and $A$ are chosen 
so that all the data points fall on a
single curve (namely $P_sN_x^{-2+\eta}=f(N_x/\xi)$) regardless of the 
system size.\cite{MS}

However, as we shall see in the following, 
these scaling methods eq.\ref{eqFSS}  turns out to suffer 
from finite size effects when 
the system size is small and/or the density of states at the 
Fermi level is small. In fact, in the former method mentioned above, 
$P_s$ for small system sizes do not coincide with those for 
large system sizes even at $T=T_0$. As for the latter method, 
$P_sN_x^{-2+\eta}$ for small system sizes deviate from those 
for larger system sizes at low temperatures.

Due to this problem, here we
identify $T_0$ as the temperature at which $P_s$ for the 
largest two system sizes coincide in the former method, 
while in the latter method, we search the values for $T_0$ and $A$ so 
that as many data points as possible fall on a single curve, 
although the data for small system sizes deviate from the 
curve at low temperatures.

\subsection{Extrapolation method}

Since the scaling method suffers from finite size effects on
some cases, we have adopted the third method, namely 
the extrapolation method.
This method has been adopted for the study of the (real) 
superconducting transition temperature for the three dimensional 
attractive Hubbard model.\cite{Sewer}
Since $P_s$ should increase sharply at $T=T_0$ in the 
thermodynamic limit, it is
reasonable to assume that $T_0=\lim_{N \rightarrow \infty}T_1^{(N)}$, 
where $T_1^{(N)}$ is the inflection point of $P_s(T)$ for the 
$N$ site system.  $T_1^{(N)}$ can be
obtained by fitting $P_s(T)$ by an appropriate function. In this study we
use a gaussian form $A_N \exp [-B_N(T-C_N)^2]+D_N$ as a fitting function, where
$A_N$-$D_N$ are fitting parameters.
Here, we find that $T_1^{(N)}$ scales as $1/N$ in the range of $N$ considered
in the present study. We may then obtain $T_0$ by plotting $T_1^{(N)}$ as
a function of $1/N$ and linearly extrapolating it to the $1/N \rightarrow
0$ limit.

\section{Results}
\subsection{Finite size scaling}

\begin{figure}[h]
 \begin{center}
  \includegraphics[scale=0.8]{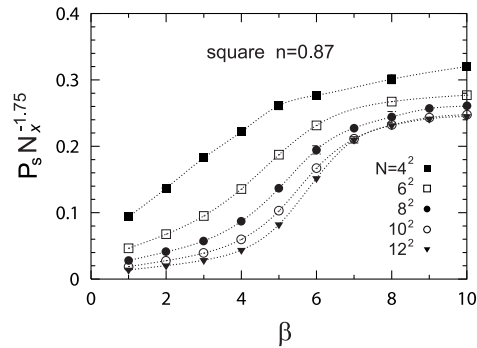}
  \caption{\small \label{sqA} $P_s N_x^{-1.75}$ plotted against $\beta$
  for the square lattice. Dashed lines are guides to the eyes.}
 \end{center}
\end{figure}
\begin{figure}[h]
 \begin{center}
  \includegraphics[scale=0.8]{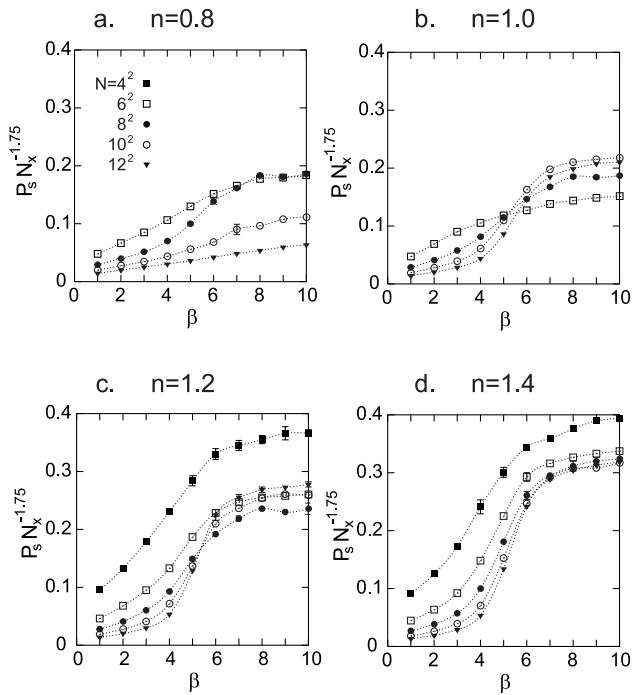}
  \caption{\small \label{triA} A plot similar to Fig.\ref{sqA} but for
  triangular lattice with several band filling. Band filling $n$ is
  $0.8$(a),$1.0$(b),$1.2$(c),$1.4$(d).}
 \end{center}
\end{figure}

We first present finite size scaling analysis,
where we identify the temperature at which $P_s N_x^{-1.75}$ coincides 
between different system sizes. 
Before going into the results for the triangular lattice,
we show the result for the square lattice with $n=0.87$,
where a scaling analysis has been performed for system sizes 
up to $8\times 8$ in ref.\onlinecite{MS}.
In Fig.\ref{sqA}, $P_s N_x^{-1.75}$ is plotted as 
functions of $\beta$ for system sizes from $4\times 4$ to $12\times 12$.
We can see that the results for the largest two system sizes, 
$10\times 10$ and $12\times 12$ merges at around $\beta=7$.
The results 
for smaller systems do not cross or merge with each other within the
present temperature range, but if we assume that the  
results for small systems are strongly affected by finite size effects,
we may adopt $T_0/t \simeq 1/7 \simeq 0.13$ for this band filling,
which is roughly consistent with what has been concluded $(T_0/t \simeq 0.1)$ 
in ref.\onlinecite{MS}.

We now move on to the results for the triangular lattice.
In Fig.\ref{triA}, $P_s N_x^{-1.75}$ is plotted as 
functions of $\beta$. For $n=1.4$, the results 
for the largest two systems, $10\times 10$ and $12\times 12$
merges at around $\beta=6$. The results 
for smaller systems do not cross or merge with each other within the
present temperature range as in the case of the 
square lattice, but if we again assume that the  
results for small systems are strongly affected by finite size effects,
we may adopt $T_0/t \simeq 1/6=0.17$ for this band filling.
For $n=1.2$, the results for the largest three system sizes cross at 
$\beta\simeq 5.3$. Again assuming that results for smaller system 
sizes are affected by finite size effects, we may take $T_0/t \simeq 0.19$
for this band filling.
As for $n=1.0$, although the results for $12\times 12$ and $8\times 8$
crosses at around $\beta=6$, those of the largest two sizes do not 
cross with each other. Therefore, $T_0/t$ cannot be estimated from 
these results at this band filling.
The situation is even worse in the case of $n=0.8$, where 
$P_s N_x^{-1.75}$ of the largest two system sizes do not 
even come close at low temperatures. Here again we cannot 
evaluate $T_0/t$ for this band filling from these results.

\subsection{Extrapolation method}
\begin{figure}[h]
\begin{center}
 \includegraphics[scale=0.8]{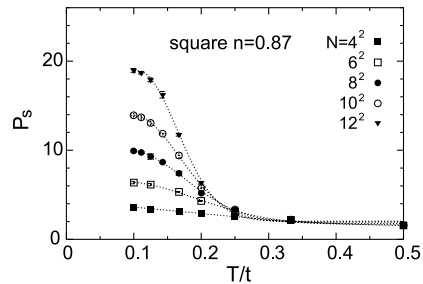}
 \caption{\small \label{rawPssq}The AFQMC data of pairing correlation
 function $P_s$ for square lattice with $n=0.87$. Dashed curves are the
 fitting results.}
\end{center} 
\end{figure}

\begin{figure}[h]
\begin{center}
 \includegraphics[scale=0.8]{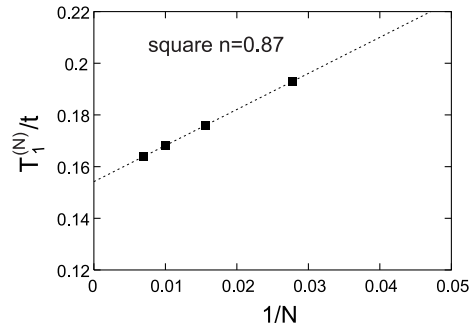}
 \caption{\small \label{sqB} $T_1^{N}$ plotted against $1/N$ for
 the square lattice with $n=0.87$. The dashed line is a least squares fit.}
\end{center} 
\end{figure}


We now evaluate $T_0$ using an alternative method, that is, 
the extrapolation method.
Here again, we first show the result for the square lattice 
with $n=0.87$. In Fig.\ref{rawPssq},
the raw $P_s$ data is plotted as functions of 
temperature for each system size. $T_1^{(N)}$ extracted from these 
data are plotted against $1/N$ in Fig.\ref{sqB}. The results for $N=4^2$
turn out to be too much affected by finite size effects, so we have
omitted these data in the extrapolating process. $T_0$ obtained by
extrapolating the results to $1/N \rightarrow 0$ is $T_0/t \simeq 0.15$,
which is a little bit higher than, but in fair agreement with the value
obtained by the scaling method.

\begin{figure}[h]
\begin{center}
 \includegraphics[scale=0.8]{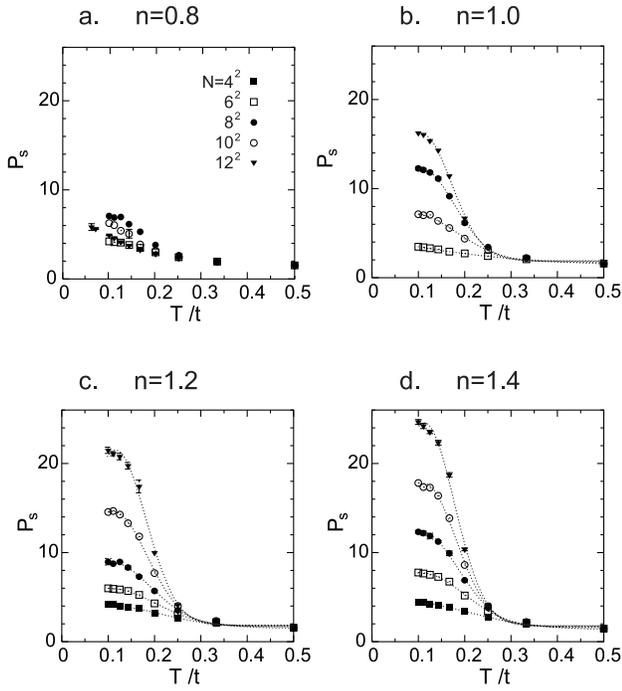}
 \caption{\small \label{rawPstri} A plot similar to Fig.\ref{rawPssq}
 but for the triangular lattice  with  $n=0.8$(a), $1.0$(b), $1.2$(c),
 $1.4$(d). Dashed lines are fitting results.}
\end{center} 
\end{figure}
\begin{figure}[h]
\begin{center}
 \includegraphics[scale=0.8]{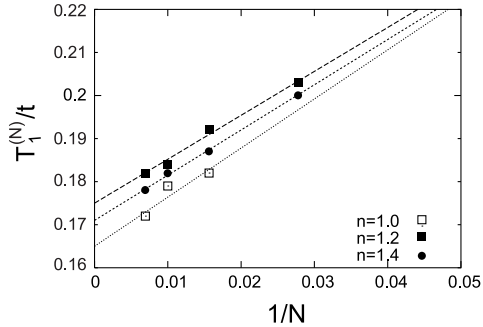}
 \caption{\small \label{triB} A plot similar to Fig.\ref{sqB} but for
 the triangular lattice with $n=1.0$, $1.2$, $1.4$.}
\end{center} 
\end{figure}
\begin{figure}[h]
\begin{center}
 \includegraphics[scale=0.8]{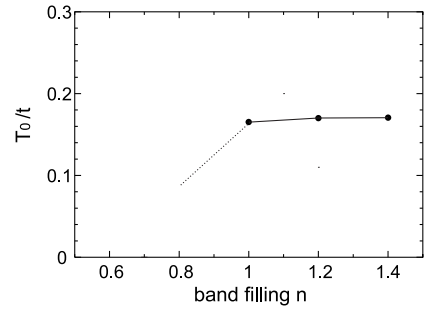}
 \caption{\small \label{T0-n} The band filling dependence of the KT
 transition temperature $T_0$. Solid and dashed lines are guides to the eyes.}
\end{center} 
\end{figure}

We move on to the triangular lattice.
In Fig.\ref{rawPstri}, the raw $P_s$ data as well as the fitting curves
are plotted as functions of $T$ for each system size and band
filling. (We do not show the fitting curves for $n=0.8$ since $T_1^{(N)}$
turns out to be negative.) $T_1^{(N)}$ obtained fromthese data are
plotted against $1/N$ for each band filling in Fig.\ref{triB}. 
$T_0$ obtained by extrapolation is $0.17$, $0.18$, $0.17$
for $n=1.4$, $n=1.2$, $n=1.0$, respectively.
For $n=1.4$ and $n=1.2$, $T_0$ obtained using the scaling method 
coincides fairly well with the values obtained here.
For $n=0.8$, $P_s$ turns out to {\it decrease} upon increasing 
the system size from $8\times 8$ to $12\times 12$, which seems to 
indicate that there is no symptom of KT transition within the 
temperature range $(>0.1t)$ investigated in the present study, 
so $T_0$, if any, should be lower than $0.1t$.
In Fig.\ref{T0-n}, $T_0$ obtained by the present method is 
plotted against the band filling.

\subsection{Justification by scaling}
\begin{figure}[h]
\begin{center}
 \includegraphics[scale=0.8]{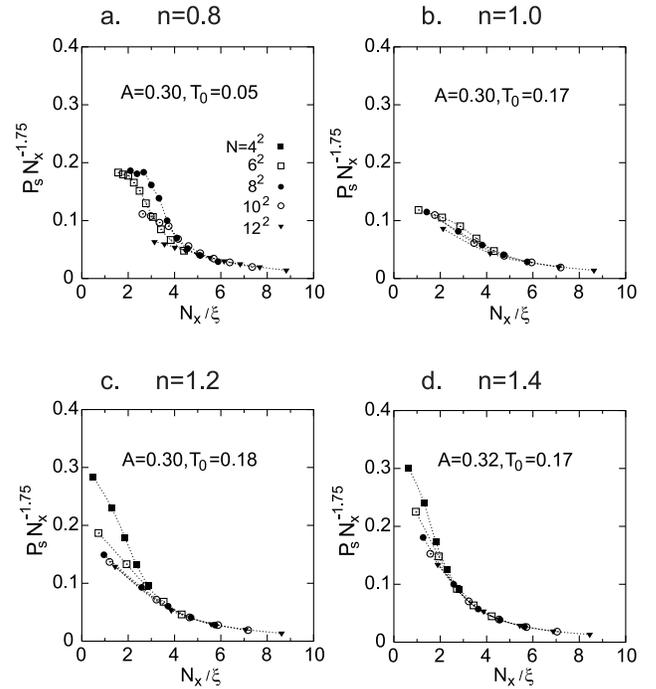}
 \caption{\small \label{triC} $P_s N_x^{-1.75}$ plotted against
 $N_x/\xi$ for the triangular lattice. }
\end{center} 
\end{figure}

\begin{figure}[h]
\begin{center}
 \includegraphics[scale=0.8]{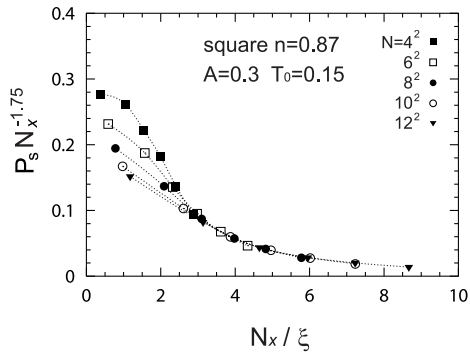}
 \caption{\small \label{sqC} A plot similar to Fig.\ref{triC} for
 the square lattice with $n=0.87$. }
\end{center} 
\end{figure}

The KT transition temperature determined by the above method can be  
further justified by checking whether $P_s$ actually 
scales as $P_s=N_x^{1.75}f(N_x/\xi)$ with $\xi=\exp(A/(T-T_0)^{1/2}))$.
In Fig.\ref{triC}, we plot $P_sN_x^{-1.75}$ as functions of 
$N_x/\xi$ for all the system sizes.
We find for $n=1.0$, $1.2$, and $1.4$ 
that the numerical results for 
$8\times 8$, $10\times 10$ and $12\times 12$ fall on 
a single curve by adopting the $T_0$ obtained 
above and choosing appropriate values of $A$.
The results for smaller systems also fall on the 
same curves at high temperatures (namely for large $N_x/\xi$), but 
at low temperatures they start to deviate due to finite size effects.
As for $n=0.8$, if $A=0.3$ and $T_0=0.05$ are chosen, the
results for $12 \times 12$ and $10 \times 10$ fall on a similar curve.
These results further justify the values of $T_0$ obtained 
in the preceding sections. Similar results are also obtained for the 
square lattice as seen in Fig.\ref{sqC}.

\section{Discussion}
\subsection{Comparison of the methods for determining $T_0$}
Our results show that the scaling method using the crossing point 
of $P_s N_x^{-1.75}$ is strongly affected by 
finite size effects, and that it is difficult to know from the 
beginning the system size necessary to obtain an accurate $T_0$.
The present analysis suggests that this method seems to work well 
at band fillings where the density of states at $E_F$ is relatively large,  
namely at $n=1.2$ and $n=1.4$ in the present case. There, the 
agreement with the results of the extrapolation method is also 
good. This may be because the discreteness of the energy 
levels due to the finite system size is small when the density of states
is large.

\subsection{Correlation between $T_0$ and the density of states}

\begin{figure}[h]
\begin{center}
 \includegraphics[scale=0.8]{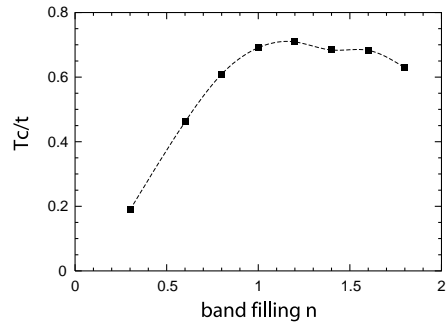}
 \caption{\small \label{MF} The band filling dependence of the $T_c^{MF}$ for
 the triangular lattice obtained by applying BCS mean field approximation. }
\end{center} 
\end{figure}

In order to look into a possible correlation between $T_0$ and the 
density of states at the Fermi level, 
we compare the present results of $T_0$ with the transition
temperature obtained by mean field approximation (Fig.\ref{MF}). Although the values of $T_c^{\rm MF}$
themselves are much larger than $T_0$, we find a similar band filling
dependence, namely, $T_c^{\rm MF}$ is almost constant for $1< n <1.5$,
and smaller for $n<1$. The origin of this similarity is not clear, 
but this does seem to suggest that $T_0$
is roughly correlated with the density of states near the Fermi level
at least for the value of $U(=-4t)$ adopted in the present study.
It would be an interesting future study to analyze this point 
from the viewpoint of the study by Timm {\it et al}, where 
the relation between the mean field $T_c$ 
and the KT superconducting transition temperature in the 
{\it repulsive} Hubbard model has been investigated by 
calculating the superfluid density as a function of temperature 
and combining it with the Berezinskii-Kosterlitz-Thouless theory.\cite{Timm}

\begin{figure}[h]
\begin{center}
 \includegraphics[scale=0.8]{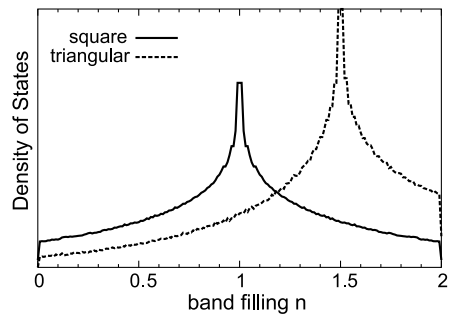}
 \caption{\small \label{dos} The density of states of the square and
 the triangular lattices on the tight binding model.}
\end{center} 
\end{figure}

$T_0$ for $1\leq n\leq 1.4$ is still larger than $T_0$ for the square lattice at
$n=0.87$ despite the fact that the density of states near the Fermi level
for the square lattice with $n=0.87$ is larger than for the triangular
lattice with $n=1.0$ (see Fig.\ref{dos}). 


If $T_0$ is indeed positively correlated with the density of states near
the Fermi level, this ``inversion'' of $T_0$ between the square and
the triangular lattice may be due to the absence of charge density
wave ordering in the latter lattice due to frustration, as discussed
previously.\cite{Santos}

\section{Conclusions}
In the present study, we have investigated the 
superconducting KT transition of the attractive Hubbard model on a
two-dimensional triangular lattice using auxiliary field quantum Monte
Carlo method for system sizes up to $12\times 12$ sites. Combining
three methods to analyze the numerical data for the pairing
correlation function, we find that the transition temperature stays almost constant within the band filling range of $1.0 \leq n \leq 1.4$,
while it is found to be much lower in the $n<1$ region.
Among the three methods, the extrapolation method is found to work
well regardless of the band filling, while the methods relying on finite
size scaling require some caution when the density of states near the
Fermi level is small and the calculation is restricted to small system
sizes.

\section{Acknowledgements}
We thank Yoichi Yanase for valuable discussions. Part of the numerical
calculations has been done at the Computer Center, ISSP, University of
Tokyo.

\end{document}